\documentclass[aps,prl,twocolumn,superscriptaddress]{revtex4}
\usepackage{epsfig,graphicx}
\usepackage{indentfirst}
\usepackage{amsfonts,amssymb,latexsym,amsmath,enumerate,amsthm}
\usepackage{bm}
\usepackage{color}

\newcommand{\be}{\begin{equation}}
\newcommand{\ee}{\end{equation}}
\newcommand{\bea}{\begin{eqnarray}}
\newcommand{\eea}{\end{eqnarray}}

\begin{document}

\title{Possibility to probe negative values of a Wigner function in scattering \\ of a coherent superposition of electronic wavepackets by atoms}

\author{Dmitry~V.~Karlovets}
\affiliation{Tomsk State University, Lenina 36, 634050 Tomsk, Russia}

\author{Valeriy~G.~Serbo}
\affiliation{Novosibirsk State University, Pirogova 2, RUS--630090, Novosibirsk, Russia}
\affiliation{Sobolev Institute of Mathematics, Koptyuga 4, RUS-630090, Novosibirsk, Russia}

\date{\today}

\begin{abstract}
Within a plane-wave approximation in scattering, an incoming wave packet's Wigner function stays everywhere positive, which obscures such purely quantum phenomena as non-locality and entanglement. 
With the advent of the electron microscopes with subnanometer-sized beams one can enter a genuinely quantum regime where the latter effects become only moderately attenuated.
Here we show how to probe negative values of the Wigner function in scattering of a coherent superposition of two Gaussian packets with a non-vanishing impact-parameter between them (a Schr\"odinger's cat state) by atomic targets. For hydrogen in the ground $1s$ state, a small parameter of the problem, a ratio $a/\sigma_{\perp}$ of the Bohr radius $a$ to the beam width $\sigma_{\perp}$, is no longer vanishing. 
We predict an azimuthal asymmetry of the scattered electrons, which is found to be up to $10 \%$ and argue that it can be reliably detected. Production of beams with the not-everywhere positive Wigner functions and probing such quantum effects can open new perspectives for non-invasive electron microscopy, quantum tomography, particle physics, etc. 
\end{abstract}


\maketitle

\textit{Introduction.} --- The conventional scattering theory deals with the wave packets whose width in momentum space is vanishingly small and their spatial extent is much larger than that of a target.
In high-energy physics there are very few examples in which this plane-wave approximation fails to work, see \cite{Hall, Impact, t2, Dodonov, Akhmedov_09, Akhmedov_10, EPL, JHEP}, 
while low-energy scattering with subnanometer-sized beams of the modern electron microscopes can be described within a paraxial (akin to WKB) framework \cite{PRA2015, PRA2017, JHEP}. 
Neglect of the transverse coherence effects led to significant discrepancies between the theory and experiments in ion-atom collisions, resolved by using the Gaussian packets \cite{Sarkadi, Sch}.
What is less realized is that the incoming wave packets' density matrices in phase space, the Wigner functions \cite{Wigner}, turn out to be everywhere positive in the paraxial regime, 
whereas their negative values may become prominent only when the beams are focused to a spot comparable to the Compton wave length \cite{JHEP}, $\sigma_{\perp} \sim \lambda_c = \hbar/(mc)$, 
which is $3.9\times 10^{-11}\, \text{cm}$ for electron. That is why effects induced by the Wigner function's regions of negativity have never been observed in particle or atomic collisions.

However, when the projectile's wave function is either not Gaussian or it represents a coherent superposition of packets, the genuine quantum effects arising from the Wigner function's possible negativity, such as non-locality and entanglement, may become noticeable at much larger scales. A simplest example here is a superposition of two coherent states separated by an impact parameter $2{\bm r}_0$, $|{\bm r}_0\rangle \pm |-{\bm r}_0\rangle$, the so-called  Schr\"odinger's cat state widely used in quantum optics where the negative values of its Wigner function are measured \cite{Ourjoumtsev, Laiho, Douce} and even teleported \cite{Lee}. Other examples of the systems with the not-everywhere positive Wigner functions embrace ensembles of thousands of cold entangled atoms \cite{McConnell}, quantum bosonic gases \cite{Fischer}, twisted photons \cite{Mirhosseini}, etc. In electron microscopy with atomic-scale resolution, non-invasive imaging, say of biological systems, requires use of the beams coherently split into two parts \cite{Kruit}. Due to the quantum interference, such beams can also possess Wigner functions that turn negative in some regions of the phase space and, therefore, their scattering cannot be described within the existing theory. 

In this Letter we show that elastic scattering of an electronic cat state by atomic targets reveals interference effects, inaccessible in the paraxial approximation. A small parameter of the problem turns out to be a ratio of the potential radius $a$ to the beam width $\sigma_{\perp}$, which is $1/\alpha = 137$ times larger than $\lambda_c/\sigma_{\perp}$ for hydrogen ($a\approx 0.053$ nm is a Bohr radius). As beams of the electron microscopes have already been focused to an {\AA}ngstr\"om size spot \cite{Angstrom, Pohl, Review}, one can enter a  genuinely quantum (or deeply non-plane-wave) regime of scattering, in which the projectile's probability density becomes spatially inhomogeneous on the atomic scale, the packets' width in momentum space is finite, and the Wigner functions are treated non-perturbatively (cf. \cite{JHEP}). Effects of the possible non-classicality are only moderately attenuated here, which is of crucial importance for the interaction-free quantum measurements \cite{Kruit, Elitzur}, for instance, and may lead to new schemes of the quantum tomography of electrons and other massive particles.

As a hallmark of strong quantum interference, we predict an azimuthal asymmetry of the scattered electrons, which reaches the values of $10\%$ when $\sigma_{\perp} \gtrsim a$ and oscillates with $r_0$ when $r_0 \gtrsim \sigma_{\perp}$. The effect does not appear within the Born approximation with the Gaussian beams, even for an off-axis collision. Moreover, the asymmetry vanishes for wide beams, $a/\sigma_{\perp} \ll 1$, when the impact parameter between the two Gaussians is large, $r_0 \gg \sigma_{\perp}$, and when the beam is just an incoherent statistical mixture of packets. We argue that this phenomenon can be reliably detected with existing technology. Beams of massive particles with the not-everywhere positive Wigner functions can become a useful tool for imaging atoms, molecules, surface inhomogeneities, biological samples, etc., along with the vortex electrons \cite{Review} or neutrons \cite{neu} for instance, for quantum information, quantum tomography, and even for particle physics. For the sake of brevity, we set $\hbar = 1$. All vectors, except ${\bm Q}$ and ${\bm p}_f$, have two components.

\textit{Wigner function of a Schr\"odinger's cat state.} --- Consider a system of two Gaussian packets moving along the $z$ axis with the same mean momentum $
\{0,0,p_i\}$, coordinate uncertainties $\sigma_z$ and $\sigma_{\perp}$, 
and separated by a distance $2{\bm r}_0 \equiv 2\{x_0,y_0,0\}$. Its transverse wave function in the momentum representation
is a coherent superposition of two packets:

\begin{eqnarray}
& \displaystyle \psi_{1\pm 1} ({\bm p}) = \frac{\psi_1 ({\bm p})}{\sqrt{2}} \frac{e^{-i {\bm r}_0\cdot {\bm p}} \pm e^{i {\bm r}_0\cdot {\bm p}}}{\sqrt{1 \pm \exp\{-{\bm r}_0^2/(2\sigma_{\perp}^2)\}}},
\cr & \displaystyle \int d^2p\, |\psi_{1\pm 1} ({\bm p})|^2 = 1,
\label{WaveF}
\end{eqnarray}
where $\psi_1 ({\bm p}) = \sqrt{2\sigma_{\perp}^2/\pi}\, \exp\left\{-\sigma_{\perp}^2 {\bm p}^2\right\}, \int d^2p\, |\psi_{1} ({\bm p})|^2 = 1,$
is a wave function of one packet. Drawing an analogy to quantum optics, we shall call the states $\psi_{1+1}$ and $\psi_{1-1}$ \textit{an even cat state} and \textit{an odd one}, respectively.
The Wigner function at $t = 0$ (see, for instance, \cite{Case}) is
\begin{eqnarray}
& \displaystyle W_{1\pm 1}({\bm r}, {\bm p}) = \int \frac{d^2k}{(2\pi)^2}\,\, e^{i{\bm k}\cdot{\bm r}}\, \psi_{1\pm 1}^* ({\bm p} - {\bm k}/2) 
\cr & \displaystyle \times \psi_{1\pm 1} ({\bm p} + {\bm k}/2) = \frac{W_1({\bm r}, {\bm p})}{1 \pm \exp\{-{\bm r}_0^2/(2\sigma_{\perp}^2)\}}
\cr & \displaystyle \times \Big (\cosh ({\bm r}_0\cdot{\bm r}/\sigma_{\perp}^2)\, e^{-{\bm r}_0^2/(2\sigma_{\perp}^2)} \pm \cos (2 {\bm r}_0\cdot{\bm p})\Big),
\label{WignF}
\end{eqnarray}
where
\begin{eqnarray}
& \displaystyle W_1({\bm r}, {\bm p}) = \frac{1}{\pi^2} \exp \left\{-2\sigma_{\perp}^2{\bm p}^2 - {\bm r}^2/(2\sigma_{\perp}^2)\right\}
\label{WignF1}
\end{eqnarray}
is an everywhere-positive Wigner function of one packet.

\begin{figure}[t]
	\center
	\includegraphics[width=1.00\linewidth]{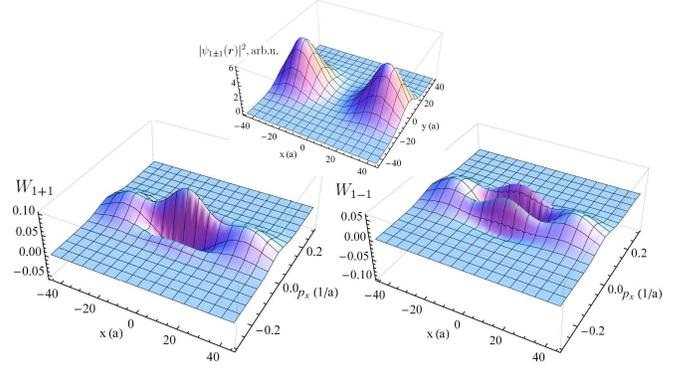}
	\caption{Wigner functions of the electronic Schr\"odinger cat state (\ref{WignF}) with $x_0 = 3\, \sigma_{\perp}$ and $\sigma_{\perp} = 10 a$. Left: an even cat state, right: an odd cat state.
	The upper panel shows the probability density, $|\psi_{1\pm 1} ({\bm r})|^2 = \int d^2p\, W_{1\pm 1}$. Parameters: $y=y_0=p_y=0,\, p_z = p_i = 10/a\, (\varepsilon_{\text{kin}} = 1.4\, \text{keV})$.
\label{Fig1}}
\end{figure}

\begin{figure}[t]
	\center
	\includegraphics[width=0.99\linewidth]{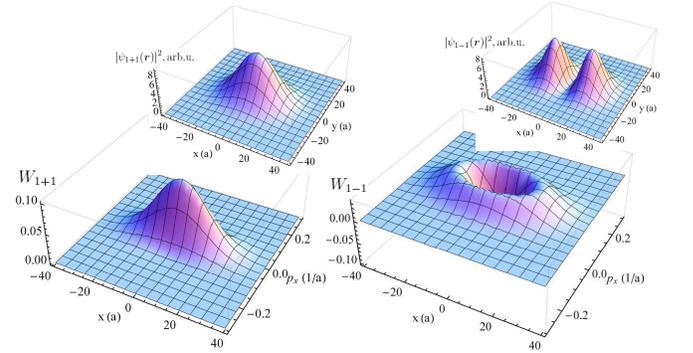}
	\caption{The same as in Fig.\ref{Fig1} but for $x_0 = \sigma_{\perp}$.
\label{Fig2}}
\end{figure}

It is the interference term, $\cos (2 {\bm r}_0\cdot{\bm p})$, that is responsible for possible negativity in some regions of the phase space.
The function $W_{1+1}$ can be negative when
\begin{eqnarray}
& \displaystyle
r_0 \gtrsim \sigma_{\perp} \sim 1/p_{\perp},
\label{negcond}
\end{eqnarray}
i.e. the distance between the two packets gets larger than the packet's width $\sigma_{\perp}$ (see Fig.\ref{Fig1}),
but stays everywhere positive when $r_0 \lesssim \sigma_{\perp}$ (see Fig.\ref{Fig2}).
The function $W_{1-1}$, on the contrary, has prominent negative regions even when $r_0 \approx \sigma_{\perp}$.
For smaller values, $r_0 <  \sigma_{\perp}$, the odd cat state, unlike the even one,
reveals a somewhat fermionic behavior with the Gaussians staying separated by the distance $\approx 2\sigma_{\perp}$, as shown in Fig.\ref{Fig2}.
That is why in what follows we stick to the case $r_0 \geq \sigma_{\perp}$ for the odd cat.

For an \textit{incoherent} mixture of two one-particle Wigner functions,
\begin{eqnarray}
\displaystyle \frac{1}{2}\,\Big(W_1({\bm r}, {\bm p};{\bm r}_0) + W_1({\bm r}, {\bm p};-{\bm r}_0)\Big ) \propto \cosh ({\bm r}_0\cdot{\bm r}/\sigma_{\perp}^2),
\label{Incoh}
\end{eqnarray}
which takes place for mixed states for instance \cite{Case}, the interference term is absent.
As a result, this sum stays everywhere positive and looks exactly like the probability density shown in the upper panel of Fig.\ref{Fig1}.

\textit{Scattering of a Schr\"odinger's cat by atoms.} --- Let us now study elastic scattering of such a system off a central potential with an effective radius $a$ in the Born approximation.
In order to describe collisions beyond the plane-wave model, one should employ the theory recently developed by us in Refs.\cite{PRA2015, PRA2017}.
We consider an experimentally relevant scenario in which (i) the packet's longitudinal size $\sigma_z$ is larger than the field radius, $\sigma_z \gg a$,
and (ii) the packet's dispersion in the transverse plane is negligible during the collision, that is, $t_{\text{dis}} \sim \sigma_{\perp}/v_{\perp} \gg t_{\text{col}} \sim \sigma_z/v_z$.
Under these assumptions, which can be rewritten as
\begin{eqnarray}
& \displaystyle
a \ll \sigma_z \ll \sigma_{\perp}^2\, p_i,
\label{Ineq}
\end{eqnarray}
the scattering amplitude is given by Eqs.\,(29),\,(30) of \cite{PRA2015} where one can put $\theta_k = 0$, as $\theta_k \approx 1/(\sigma_{\perp}p_i) \ll 1$.

Then instead of a single scatterer we take a macroscopic target with a density of atoms $n ({\bm b})$, normalized as $\int d^2b\, n ({\bm b}) = 1$, and with a width $\sigma_t \gg a$.
A number of scattering events for an incident beam of $N_e$ electrons, according to Eq.(41) from \cite{PRA2017}, is
\begin{eqnarray}
& \displaystyle \frac{d\nu}{d\Omega} = N_e \int d^2 b\, n ({\bm b})\, \frac{d^2 p}{2 \pi}\frac{d^2 k}{2 \pi}\, e^{i{\bm b}\cdot{\bm k}}\, f({\bm Q} - {\bm p} - {\bm k}/2) 
\cr & \displaystyle \times f^*({\bm Q} - {\bm p} + {\bm k}/2) \psi ({\bm p} + {\bm k}/2) \psi^* ({\bm p} - {\bm k}/2),
\label{dN}
\end{eqnarray}
where we have changed the variables: ${\bm k}_{\perp}, {\bm k}_{\perp}^{\prime} \rightarrow {\bm p} = ({\bm k}_{\perp}+{\bm k}_{\perp}^{\prime})/2, {\bm k} = {\bm k}_{\perp}-{\bm k}_{\perp}^{\prime}$,
and ${\bm Q} = {\bm p}_f - p_i \hat{{\bm z}}$ with ${\bm p}_f$ being a $3$D momentum of the final (plane-wave) electron:
$$
Q_z = p_f \cos \theta - p_i,\ {\bm Q}_{\perp} = p_f \sin \theta \{\cos \varphi, \sin \varphi, 0\}.
$$
The scattering amplitude $f ({\bm Q})$ is a real function in the Born approximation.

When the target density $n ({\bm b})$ varies at distances of the order of $\sigma_t \gg a$, only the small values of $k \lesssim 1/\sigma_t \ll 1/a$ contribute to the integral (\ref{dN}).
Therefore one can expand the amplitudes as follows:
\begin{eqnarray}
\displaystyle f({\bm Q} - {\bm p} - {\bm k}/2) f^*({\bm Q} - {\bm p} + {\bm k}/2) = (f({\bm Q} - {\bm p}))^2 + 
\cr \displaystyle + \mathcal O (k^2 a^2),\ k^2 a^2 \lesssim a^2/\sigma_t^2.
\label{fexp}
\end{eqnarray}
Within this accuracy, the integral over ${\bm k}$ brings the projectile's Wigner function $W ({\bm b}, {\bm p})$ and we arrive at the following basic formula
\begin{eqnarray}
& \displaystyle \frac{d\nu}{d\Omega} = N_e \int d^2 b\, d^2 p\,\, n ({\bm b})\, W ({\bm b}, {\bm p}) (f({\bm Q} - {\bm p}))^2,
\label{dNn}
\end{eqnarray}
which couples the number of scattering events to the target's profile and to the Wigner function. 
Clearly, possible negative values of the latter diminish the number of events due to \textit{destructive self-interference}.
When the target is homogeneous and wide, one can put $n ({\bm b}) = \theta (R - b)/(\pi R^2)$, and the corresponding number of events
reproduces Eq.(38) of \cite{PRA2015}, from which the standard (plane-wave) Born results easily follow.

As a simplest example we take a Gaussian packet with the everywhere positive Wigner function (\ref{WignF1}) scattered by a Gaussian target with
\begin{eqnarray}
\displaystyle
n ({\bm b}) = \frac{1}{2\pi \sigma_t^2}\, e^{-({\bm b} - {\bm b}_0)^2/(2 \sigma_t^2)}.
\label{nt}
\end{eqnarray}
The result is
\begin{eqnarray}
\displaystyle \frac{d\nu_1}{d\Omega} = \frac{N_e \sigma_{\perp}^2}{\pi^2 \Sigma^2}\,e^{-b_0^2/(2\Sigma^2)} \int d^2 p\, (f({\bm Q} - {\bm p}))^2\, e^{-2\sigma_{\perp}^2 p^2},
\label{dN_1}
\end{eqnarray}
where $ \Sigma^2 = \sigma_t^2 + \sigma_{\perp}^2$.
This number of events depends as $\exp\{-b_0^2/(2\Sigma^2)\}$ upon the distance $b_0$ between the center of the target and that of the incident packet. 
For more sophisticated targets, say those with surface inhomogeneities, varying this distance $b_0$ while detecting scattered electrons at a certain angle
would allow one to perform imaging of the surface. On the other hand, for a given target's profile $n ({\bm b})$, measuring angular distributions of the electrons 
allows one to perform \textit{quantum tomography} of the incident electron's state (cf. \cite{Jullien}). 

It is remarkable that Eq.(\ref{dN_1}) does not depend on the azimuthal scattering angle $\varphi$, 
as one can substitute $\varphi_p - \varphi \rightarrow \varphi_p$, even if the target's center does not coincide with that of the packet (an off-axis collision).
Such azimuthal degeneracy is a consequence of the Born approximation, the condition of the wide target, and the axial symmetry of the input beam (\ref{WignF1}).
In order to restore this dependence one should either go beyond the Born approximation, or take a small non-Gaussian target, 
which necessitates taking into account the terms $\mathcal O (k^2 a^2)$ in (\ref{fexp}), or take an incident beam with no azimuthal symmetry, 
but with a strong quantum interference between the wave packets.



Indeed, for an incoherent (no interference) sum like (\ref{Incoh}) Eq.(\ref{dNn}) does not reveal azimuthal dependence, 
even if the resultant beam is not symmetric. The azimuthal dependence in the number of events $d\nu$ arises
when the in-state is either a coherent \textit{non-symmetric} beam with an everywhere positive Wigner function (say, a Gaussian with $\sigma_{\perp,x} \ne \sigma_{\perp,y}$) 
or a superposition of packets with a Wigner function that is not everywhere positive. 
In the latter (the ``most quantum'') case, we take the Schr\"odinger cat (\ref{WignF}) and the same Gaussian target (\ref{nt}). 
This time we obtain
\begin{eqnarray}
& \displaystyle \frac{d\nu_{1\pm 1}}{d\Omega} = N_e \frac{\sigma_{\perp}^2}{\pi^2 \Sigma^2} \frac{\exp\{-{\bm b}_0^2/(2\Sigma^2)\}}{1 \pm \exp\{-r_0^2/(2\sigma_{\perp}^2)\}}
\cr & \displaystyle \times \int d^2 p\, (f({\bm Q} - {\bm p}))^2\, e^{-2\sigma_{\perp}^2{\bm p}^2}
\cr
& \displaystyle \times \Big (\cosh \left ({\bm b}_0\cdot{\bm r}_0/\Sigma^2\right) e^{-r_0^2/(2\Sigma^2)} \pm \cos (2 {\bm r}_0\cdot{\bm p})\Big ),
\label{dN_app}
\end{eqnarray}
and dependence upon the azimuthal angles of ${\bm b}_0$ and ${\bm p}_f$ persists, regardless of the amplitude $f$.

As a specific example, we take a target made of hydrogen in the ground $1s$ state for which \cite{PRA2017}
\begin{eqnarray}
& \displaystyle
f ({\bm p}) = \frac{a}{2} \left(\frac{1}{1 + (a/2)^2 {\bm p}^2} + \frac{1}{(1 + (a/2)^2 {\bm p}^2)^2}\right ).
\label{Ampl}
\end{eqnarray}
Then using Eq.(54) from \cite{PRA2015} we can integrate over ${\bm p}$ and get the following final result:
\begin{eqnarray}
& \displaystyle \frac{d\nu_{1\pm 1}}{d\Omega} = \mathcal N_{1\pm 1}\, \int\limits_0^{\infty} dx\, e^{-x g}\, \frac{x + x^2 + x^3/6}{1 + x a^2/(8\sigma_{\perp}^2)}
\cr & \displaystyle \times \Bigg (\cosh \left (\frac{{\bm b}_0\cdot{\bm r}_0}{\Sigma^2}\right) e^{-r_0^2/(2\Sigma^2)} \pm
\cr
& \displaystyle \pm \cos \left(2 {\bm r}_0\cdot{\bm p}_f \frac{x a^2/(8\sigma_{\perp}^2)}{1 + x a^2/(8\sigma_{\perp}^2)} \right) 
\cr
& \displaystyle \times \exp\left\{-\frac{r_0^2}{2\sigma_{\perp}^2 (1 + x a^2/(8\sigma_{\perp}^2))}\right\}\Bigg ),
\label{dN_fin}
\end{eqnarray}
where
\begin{eqnarray}
& \displaystyle
\mathcal N_{1\pm 1} =  \frac{N_e}{2\pi\Sigma^2} \left(\frac{a}{2}\right)^2 \, \frac{e^{-b_0^2/(2\Sigma^2)}}{1 \pm e^{-r_0^2/(2\sigma_{\perp}^2)}},
\cr
& \displaystyle g = 1 + \left(\frac{a}{2}\right)^2 \left (Q_z^2 + \frac{Q_{\perp}^2}{1 + x  a^2/(8\sigma_{\perp}^2)} \right ).
\label{dN_fin_2}
\end{eqnarray}
One may call $d\sigma_{1\pm 1}/d\Omega = 2\pi\Sigma^2 N_e^{-1}\, d\nu_{1\pm 1}/d\Omega$ an effective \textit{cross section}, as it stays finite for the wide target with $\Sigma \approx \sigma_t \gg \sigma_{\perp}$.

\begin{figure}[t]
	\center
	\includegraphics[width=0.90\linewidth]{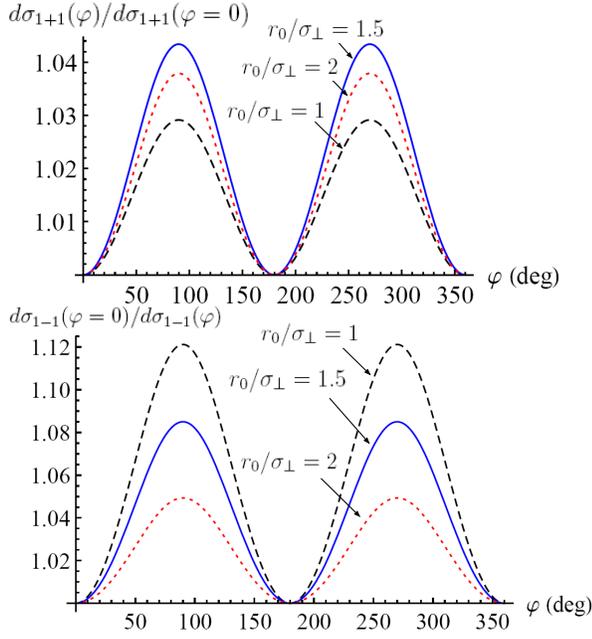}
\caption{Azimuthal asymmetry of the electrons scattered by a wide hydrogenic target for $\sigma_{\perp} = 2a \approx 0.1$ nm (FWHM $\approx 0.25$ nm), $\theta = 10^{\circ}$, $r_0/\sigma_{\perp} = 1$ (\text{black dashed line}), $1.5$ (\text{blue solid line}), $2$ (\text{red dotted line}). Parameters: $\sigma_t \gg \sigma_{\perp}, \varphi_{r_0} = 0, p_i = p_f = 10/a$ ($\varepsilon_{\text{kin}} = 1.4$ keV).
\label{Fig_3_1804}}
\end{figure}

\begin{figure}[t]
	\center
	\includegraphics[width=0.8\linewidth]{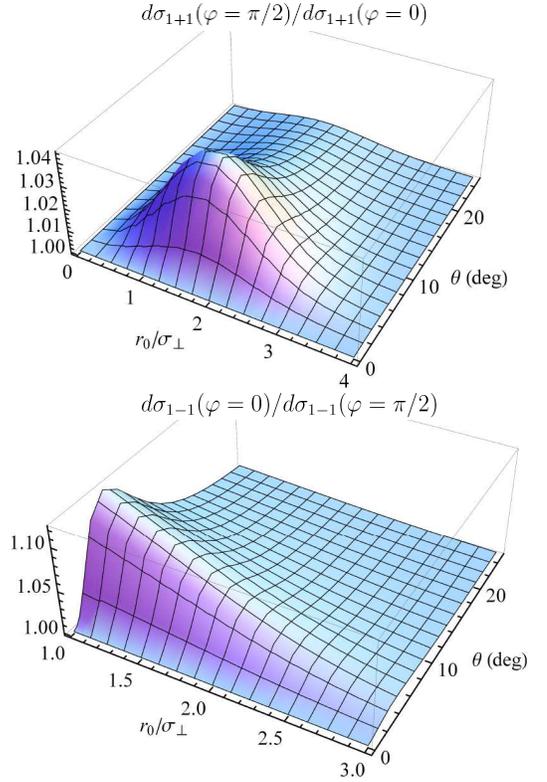}
\caption{Azimuthal asymmetry for the same parameters as in Fig.\ref{Fig_3_1804}, but with $p_i = p_f = 20/a$ ($\varepsilon_{\text{kin}} = 5.6$ keV). 
The even cat state (upper panel) reveals a periodic dependence on the impact parameter $r_0$, a hallmark of the quantum interference.
\label{Fig_4_1804}}
\end{figure}

The interference term in (\ref{dN_fin}) turns to unity for the wide packets, $a \ll \sigma_{\perp}$ (paraxial regime), but for the well-focused ones, $\sigma_{\perp} \gtrsim a$, 
destructive interference between the two parts of the beam results in the azimuthal asymmetry. This asymmetry exists only for the moderate distances between the packets,
\begin{eqnarray}
& \displaystyle
r_0 \gtrsim \sigma_{\perp} \gtrsim a,
\label{Ineq}
\end{eqnarray}
and it vanishes when either $r_0 \rightarrow 0$ (for the even cat state, for the odd one this limit has no sense) or $r_0 \gg \sigma_{\perp}$.
Two coherent beams of electrons focused to $\sigma_{\perp} \sim 0.1$ nm, as in the experiments \cite{Angstrom, Pohl}, and separated by a comparable distance can do the job.

As shown in Figs.\,\ref{Fig_3_1804},\,\ref{Fig_4_1804}, azimuthal variation of the cross section can reach $10\%$ for such beams. Clearly, the effect is stronger for the odd cat state with $r_0 \approx \sigma_{\perp}$, but for $r_0 = 2\sigma_{\perp}-3\sigma_{\perp}$ it is nearly the same as for the even cat. Width of the distributions over the scattering angle $\theta$ in Fig.\ref{Fig_4_1804} is determined by the packet's mean momentum $p_i$. For higher energies, the peak shifts to smaller angles (it's around $5^{\circ}$ for $p_i = 30/a,\, \varepsilon_{\text{kin}} = 12.5$ keV), whereas for lower ones it moves to larger polar angles and widens. Meanwhile, the value of the azimuthal asymmetry persists. With the increasing beam width $\sigma_{\perp}$, the asymmetry drops as $\sigma_{\perp}^{-2}$. For instance, for $\sigma_{\perp} = 4a$ (FWHM $\approx 0.5$ nm) it is $0.1\% - 1\%$. 

Not only does the quantum interference bring about the angular asymmetry, it also results in a periodic dependence of the latter with $r_0$, which is more pronounced for the even cat state, see the upper panel in Fig.\ref{Fig_4_1804}. Even though the asymmetry of the same order of magnitude can also originate in scattering of a non-symmetric beam with an everywhere positive Wigner function (say, when $\sigma_{\perp,x} \sim 0.1$ nm, $\sigma_{\perp,y} \gg \sigma_{\perp,x}$), the corresponding cross section would not contain the interference term, similar to Eq.(\ref{dN_1}), and hence would not oscillate with the beam width. This allows one to distinguish between the classical effects of the beam shape and the purely quantum ones evoked by the Wigner function's negativity.

\textit{Experimental feasibility.} --- Beams of the modern electron microscopes already satisfy the needed requirements. An optimal energy range is from one to several tens of keV, 
as for the energies higher than 100 keV relativistic corrections may become important and the asymmetry would have to be detected at too small polar angles. 
Coherent splitting of a beam of these energies into two parts can be achieved by employing diffraction at a periodic surface potential
or at a standing light wave due to the Kapitza-Dirac effect (see technical details and discussion of other methods in \cite{Kruit}).

When averaging over many collisions with $\sigma_{\perp} \sim 0.1 - 0.2\, \text{nm}$, one needs to control the impact-parameter $r_0$ 
with an accuracy of at least $\delta r_0 \sim 0.5\sigma_{\perp}$, and the asymmetry will be robust against smaller variations of $r_0$. 
Simultaneously, both the packets must be azimuthally symmetric. Our calculations show that a $10-20\%$ difference between $\sigma_{\perp,x}$ and $\sigma_{\perp, y}$
would result in a (false) asymmetry of $1-2\%$ at the same scattering angles $\theta$, although without the periodic dependence on $r_0$. 
The latter can be proved by making no more than three sets of measurements, say at $r_0 = \sigma_{\perp}$, $2\sigma_{\perp}$, and $3\sigma_{\perp}$. In view of the great progress made in recent years in the electron microscopy and holography with atomic resolution, such delicate control and manipulations, though challenging, seem to be within technological limits.

\textit{Summary.} --- We showed that by studying elastic scattering of the coherent superposition of electrons by atomic targets, 
one can detect a contribution of the Wigner function's negative values, which would reveal itself in the azimuthal asymmetry.
For subnanometer-sized beams we predict the asymmetry up to several percent, which, if detected, 
would be an evidence of the previously unexplored ultra-quantum regime of scattering.


Moreover, scattering of electrons by atoms may serve not only for the surface imaging with atomic resolution, 
but also as a novel method for quantum tomography of these beams. In high-energy physics, scattering of the states with the not-everywhere positive Wigner functions
can become a useful tool for probing phases of the scattering amplitudes (cf. \cite{JHEP, EPL, Dodonov}).
Along with the Schr\"odinger cats, such beams may include coherent superpositions of the vortex electrons and neutrons, of the Airy beams \cite{Airy}, 
and other non-Gaussian matter waves.


\begin{acknowledgments}
We are grateful to V.\,G.~Bagrov, I.\,F.~Ginzburg, A.\,K.~Gutakovskii, P.~Kruit, J.~Verbeeck, and to the referees for useful discussions. D.K. is supported by the Competitiveness Improvement Program of the Tomsk State University. V.G.S. is supported by the Russian Foundation for Basic Research via Grant No. 15-02-06444. D.K. also wishes to thank the A. von Humboldt Foundation (Germany). 
\end{acknowledgments}

\end{document}